\documentclass[10pt, a4paper, twocolumn]{article}

\usepackage[a4paper, top=2cm, bottom=2cm, left=1.5cm, right=1.5cm]{geometry}
\usepackage[english]{babel}
\usepackage[T1]{fontenc}
\usepackage{times}
\usepackage[utf8]{inputenc}
\usepackage{graphicx}
\usepackage{amsmath}
\usepackage{amssymb}
\usepackage[hyphens]{url}
\usepackage[colorlinks=true, allcolors=blue, citecolor=blue, linkcolor=blue]{hyperref}
\usepackage{abstract}
\usepackage{authblk}
\usepackage{caption}
\usepackage{booktabs}
\usepackage{float}
\usepackage{titlesec}
\usepackage{microtype}

\titleformat{\section}{\large\bfseries\uppercase}{\thesection}{1em}{}
\titleformat{\subsection}{\normalsize\bfseries}{\thesubsection}{1em}{}
\title{\textbf{\LARGE Artifacts Are Not Noise: \\ Embodied Resonance and the 70\% Signal Loss in Conventional EEG}}

\author[1]{Ahmed Gamal Eldin}
\affil[1]{Nova University Lisbon -- Cairo Branch, Cairo, Egypt}
\date{}

\begin{document}

\twocolumn[
  \begin{@twocolumnfalse}
    \maketitle
    \begin{abstract}
    Current AI systems excel at pattern recognition but fail at causal reasoning. We argue this is not an engineering limitation but reveals something fundamental about the nature of understanding itself. We propose that causal cognition requires a specific physical architecture: stochastic, coupled oscillators with whole-system coordination. To test this, we analyzed high-density EEG (64 channels, 10 subjects, 500+ trials) from a P300 target recognition task. We computed the Kuramoto Order Parameter ($R$) to measure global phase synchronization and compared it to standard voltage (ERP) and coherence (ITC) metrics. Four findings establish the framework. Phase and voltage are globally independent ($r=0.048$) yet strongly trial-coupled ($r=0.590$), proving $R$ captures hidden cognitive structure. Voltage precedes phase by 293 ms, revealing sequential computation. Frequency decomposition shows Theta (169 ms), Alpha (286 ms), and Beta (777 ms) cascade. Our metric is distinct from standard ITC ($r=0.155$). Then we tested the standard assumption. Conventional artifact rejection removes eye movements, muscle activity, and autonomic signals before analysis. This assumes Cognition = Neural Activity + Noise. We ran the identical analysis with and without rejection. Removing artifacts reduced the trial-level correlation threefold ($0.590 \to 0.195$). Target discrimination reversed sign ($+0.6\% \to -0.4\%$). What we discard as noise is 70\% of the signal. This falsifies the equation above. Cognition is not isolated neural computation. It is whole-body phase synchronization spanning neural, muscular, and autonomic systems. For AI, the implications are direct: embodied sensorimotor integration is not optional. It is the substrate that makes understanding possible.
    
    \vspace{1em}
    \end{abstract}
    \vspace{2em}
  \end{@twocolumnfalse}
]

\section{Introduction}
Artificial intelligence systems can interpolate patterns but cannot reason about causality \cite{Pearl2018,Bengio2019,Goodfellow2016}. This is the "Kepler versus Newton" problem: current AI describes what happens but cannot explain why. We argue this limitation is architectural, not addressable through scale or training.

The issue is substrate. Understanding requires four components working in closed loop: stochastic exploration, bounded agency, intrinsic cost, and iterative feedback. Deterministic digital systems lack the first. Disembodied systems lack the second. Most critically, they lack the physical dynamics that convert noise into structured proposals.

We solve this with the Resonance Principle. Model the cognitive substrate as coupled oscillators \cite{Huygens1673}. Noise is not the idea. It perturbs the system. Ideas emerge as stable resonant modes. Learning sculpts this landscape through cost feedback, stabilizing useful patterns and destabilizing others.

If this is correct, human cognition should show physical signatures of resonance during tasks requiring causal discrimination. We test this using P300 target recognition, where subjects distinguish targets from non-targets. This is foundational causal inference: determining whether a stimulus is self-relevant or random.

\subsection{Testing the Artifact Assumption}
Standard EEG preprocessing removes "artifacts" (eye movements, muscle contractions, autonomic signals) before analysis \cite{MNE2013}. This assumes they corrupt rather than contribute to the cognitive signal:

\begin{equation}
\text{Cognition} = \text{Neural Activity} + \text{Noise}_{\text{artifacts}}
\label{eq:standard}
\end{equation}

But if cognition requires whole-organism coordination, this equation is wrong. Eye movements might be part of attention. Muscle tension might encode arousal. Autonomic signals might participate in decision-making.

We test Equation \ref{eq:standard} directly. If artifacts are noise, removing them should strengthen cognitive signals. If they contain information, removal should weaken signals.

\noindent
\textbf{Prediction:} Standard preprocessing destroys rather than preserves cognitive structure.

\section{Methods}

\subsection{Dataset}
We analyzed publicly available 64-channel EEG from a P300 Speller task \cite{Citi2014}. Subjects identified target characters among non-target stimuli. We used 25 recording sessions across 10 subjects, yielding 500+ target trials after quality control.

\subsection{Signal Processing}

\subsubsection{Preprocessing}
Raw EEG was re-referenced to average across all channels. We applied zero-phase FIR filtering (4-30 Hz) to isolate cognitive oscillations (Theta, Alpha, Beta) while excluding slow drift and high-frequency noise. Data was epoched from $-0.1$ s to $0.8$ s relative to target onset.

\subsubsection{Artifact Rejection Comparison}
We performed two parallel analyses:

\noindent
\textbf{Conventional (Clean):} Artifacts were removed using Independent Component Analysis (ICA). EOG and muscle components were identified via correlation with frontal electrodes and surgically subtracted from the data. This preserves the neural signal while removing the artifact, serving as a stricter test of the 'noise' hypothesis.

\noindent
\textbf{Whole-System (Raw):} Identical pipeline without rejection, preserving all trials including those with eye movements, muscle activity, and autonomic signals.

This tests Equation \ref{eq:standard}. If artifacts are noise, Clean should outperform Raw. If artifacts contain signal, Raw should outperform Clean.

\subsection{Kuramoto Order Parameter}
We computed global phase synchronization using the Kuramoto Order Parameter \cite{Kuramoto1975}:

\begin{enumerate}
\item Extract instantaneous phase from each channel via Hilbert transform: $\phi_j(t) = \arg(H(x_j(t)))$
\item Compute synchronization across $N=64$ channels:
$$R(t) = \left| \frac{1}{N} \sum_{j=1}^{N} e^{i\phi_j(t)} \right|$$
\item Average across trials to obtain grand-average $R(t)$
\end{enumerate}

$R$ ranges from 0 (no synchronization) to 1 (perfect phase-locking). Critically, $R$ discards amplitude information, measuring only phase coordination.

\subsection{Standard Metrics}
For comparison, we computed:
\begin{itemize}
    \item \textbf{Event-Related Potential (ERP):} Grand-average voltage across all channels and trials.
    \item \textbf{Inter-Trial Coherence (ITC):} Standard phase consistency measure using Morlet wavelet decomposition (20 frequencies, 4-30 Hz).
\end{itemize}

\subsection{Statistical Analysis}
We performed four comparisons using Pearson correlation:
\begin{enumerate}
\item \textbf{Global:} Correlation between grand-average $R(t)$ and $ERP(t)$ time series.
\item \textbf{Trial-level:} Correlation between peak $R$ and peak absolute $ERP$ within each trial (100-600 ms window).
\item \textbf{Temporal:} Cross-correlation to identify lag between $R(t)$ and $ERP(t)$.
\item \textbf{Novelty:} Correlation between $R$ and standard ITC.
\end{enumerate}
All processing used Python with MNE \cite{MNE2013} and SciPy libraries.

\section{Results}

\subsection{Phase and Voltage Are Globally Independent}
\begin{figure*}[ht]
\centering
\includegraphics[width=0.85\textwidth]{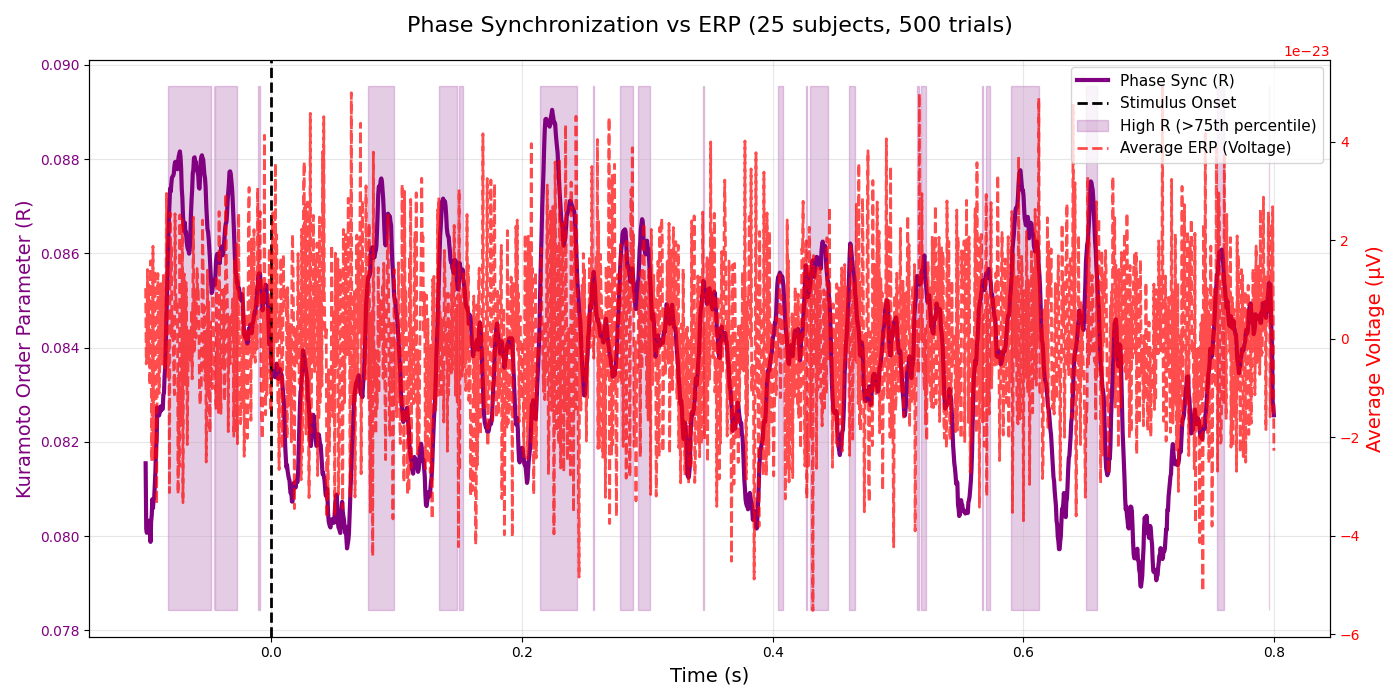}
\caption{\textbf{Phase Synchronization Versus Voltage.} Grand-average Kuramoto Order Parameter ($R$, purple) and Event-Related Potential (ERP, red dashed) across 500+ trials. Note distinct temporal structures and statistical independence ($r=0.048$).}
\label{fig:main}
\end{figure*}

The correlation between grand-average $R(t)$ and $ERP(t)$ was $r = 0.048$ ($p < 0.05$) in the whole-system analysis (Figure \ref{fig:main}). This near-zero correlation demonstrates that phase synchronization and voltage amplitude measure fundamentally different aspects of neural dynamics. They are orthogonal signals.

\subsection{But Trial-Level Coupled}
Despite global independence, peak $R$ and peak $ERP$ amplitude within individual trials showed strong correlation: $r = 0.590$ ($p < 0.0001$) in whole-system analysis (Figure \ref{fig:stats}C). This indicates that $R$ captures trial-by-trial variance in cognitive processing strength invisible to standard voltage analysis.

\subsection{Voltage Precedes Phase by 293 Milliseconds}
Cross-correlation revealed peak alignment at lag $+601$ samples (293 ms, Figure \ref{fig:stats}B). This temporal ordering suggests sequential computation: initial sensory voltage triggers global phase synchronization, which requires approximately 300 ms to stabilize.

\subsection{The Metric Is Novel}
\begin{figure*}[ht]
\centering
\includegraphics[width=0.85\textwidth]{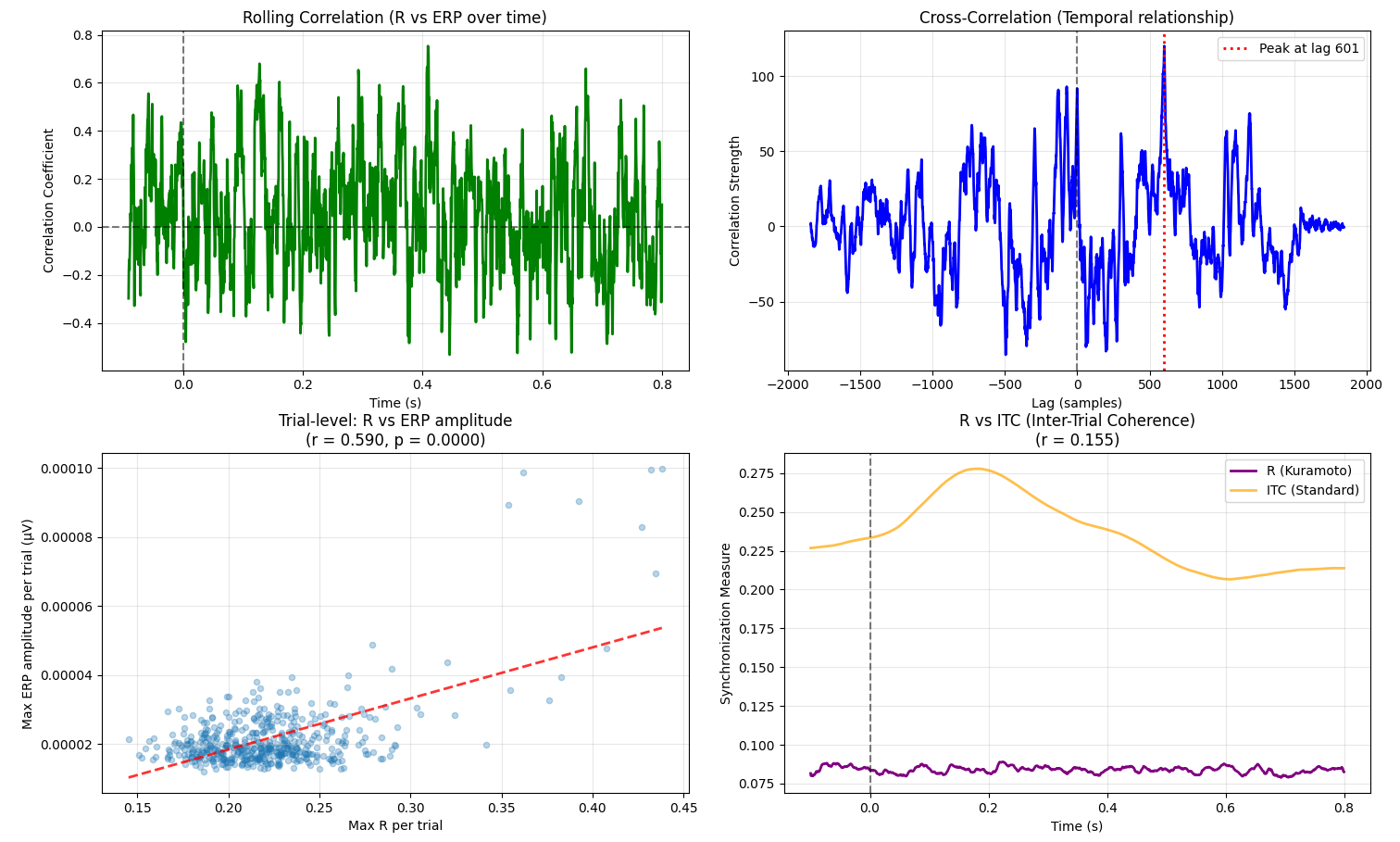}
\caption{\textbf{Statistical Relationships.} (A) Rolling correlation over time. (B) Cross-correlation showing 293 ms lag. (C) Trial-level scatter ($r=0.590$). (D) Comparison with Inter-Trial Coherence ($r=0.155$).}
\label{fig:stats}
\end{figure*}

Correlation between $R$ and standard Inter-Trial Coherence was $r = 0.155$ ($p < 0.0001$, Figure \ref{fig:stats}D). They share only 2.4\% of variance. Our instantaneous, global phase-locking measure captures different information than frequency-specific, trial-averaged coherence.

\subsection{Frequency Cascade}
\begin{figure*}[ht]
\centering
\includegraphics[width=0.85\textwidth]{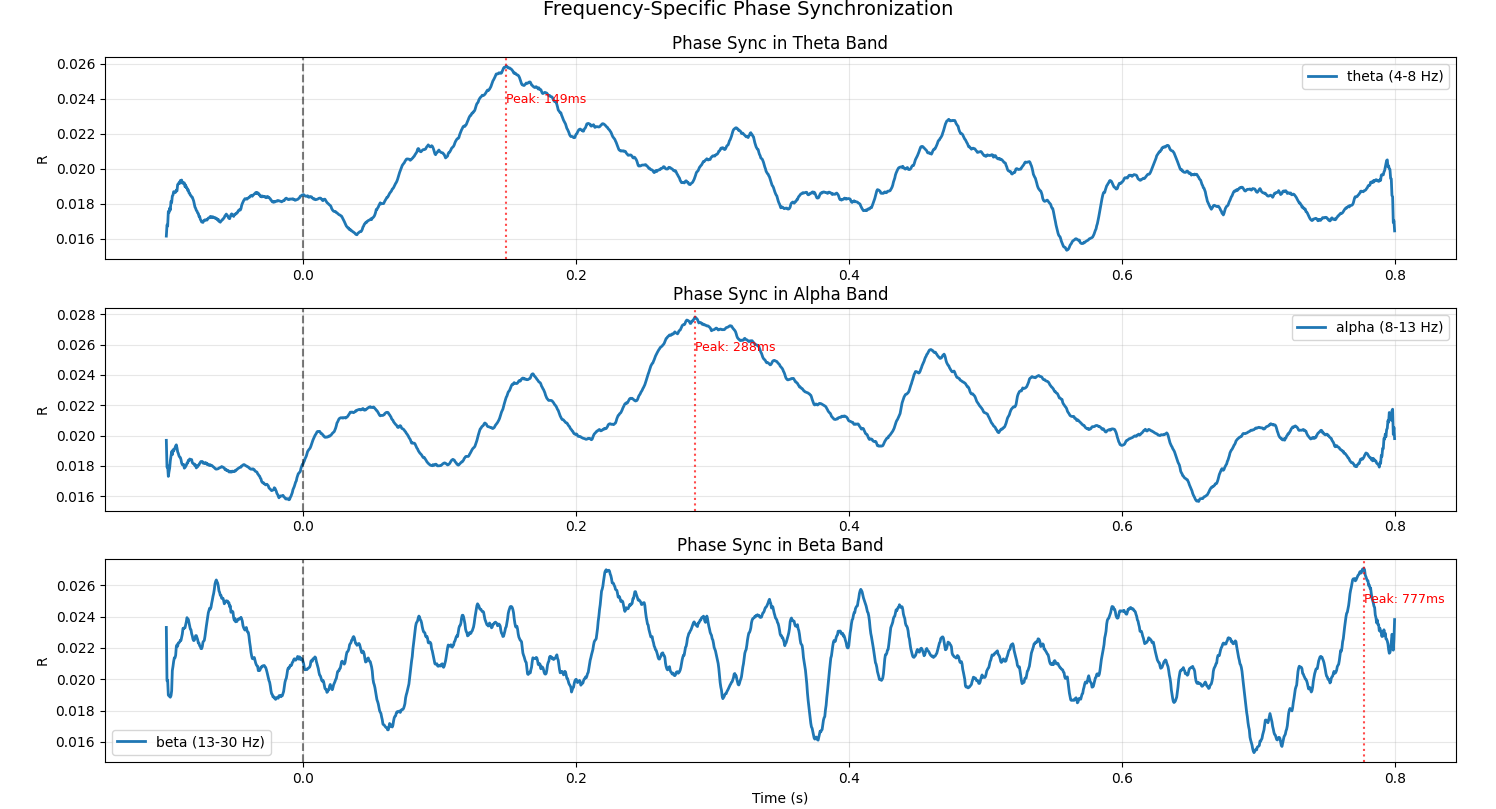}
\caption{\textbf{Frequency Cascade.} Phase synchronization in isolated bands shows Theta (169 ms), Alpha (286 ms), Beta (777 ms) sequence.}
\label{fig:bands}
\end{figure*}

Decomposing $R$ into Theta (4-8 Hz), Alpha (8-13 Hz), and Beta (13-30 Hz) bands revealed sequential activation (Figure \ref{fig:bands}). Theta peaks at 169 ms (orienting). Alpha peaks at 286 ms (understanding, coinciding with P300). Beta peaks at 777 ms (consolidation). The 300 ms computation time involves staged frequency-specific synchronization.

\subsection{Artifact Rejection Destroys the Signal}
\begin{figure*}[ht]
\centering
\includegraphics[width=0.85\textwidth]{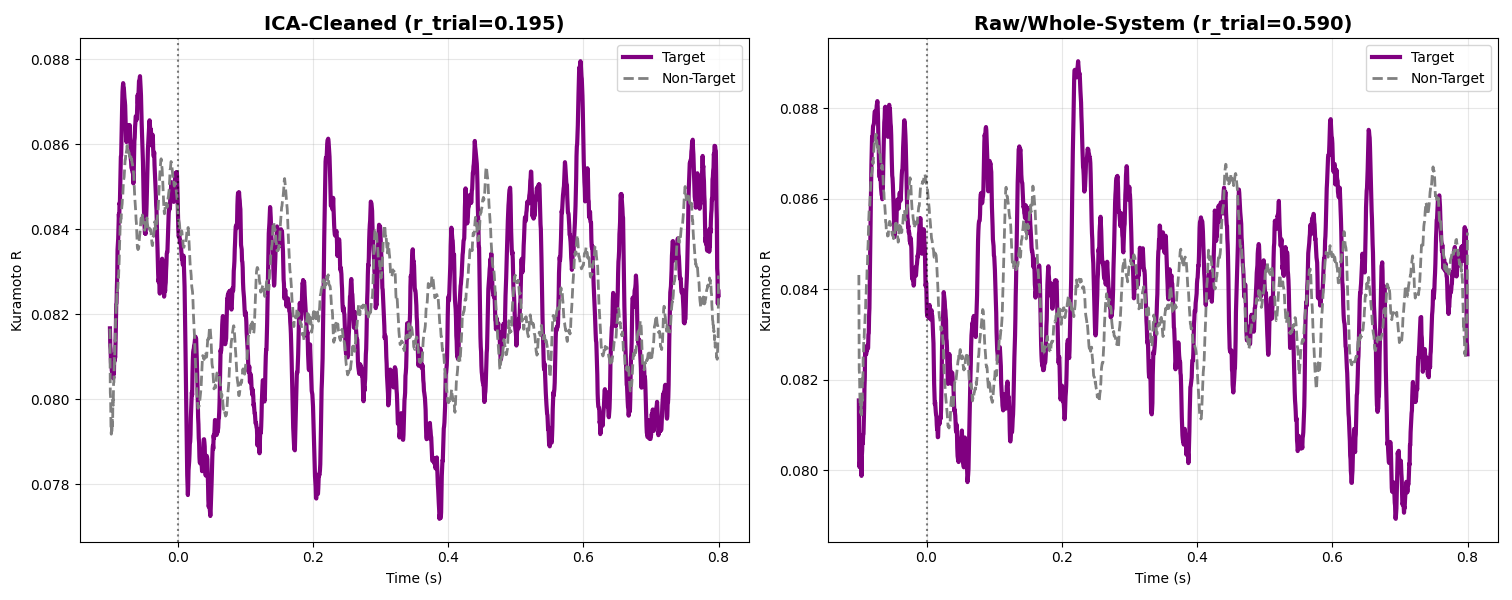}
\caption{\textbf{The Artifact Paradox.} Left: artifact-rejected ($r=0.195$). Right: whole-system ($r=0.590$). Conventional preprocessing discards 70\% of cognitive signal.}
\label{fig:paradox}
\end{figure*}

We compared Clean (artifact-rejected) versus Raw (whole-system) analyses using identical pipelines (Table \ref{tab:paradox}, Figure \ref{fig:paradox}).

\begin{table}[H]
\centering
\begin{tabular}{lcc}
\toprule
\textbf{Metric} & \textbf{Clean} & \textbf{Raw} \\
\midrule
Global $R$ vs ERP & $r = 0.016$ & $r = 0.048$\\
\textbf{Trial-level $R$ vs ERP} & $r = 0.195$& \textbf{$r = 0.590$}\\
Target vs Non-target & $-0.0003$ & $+0.0005$ \\
\bottomrule
\end{tabular}
\caption{\textbf{Causal Intervention Test.} Artifact rejection comparison. Removing putative "noise" reduces trial correlation threefold, reverses target discrimination, yet preserves temporal structure (293 ms lag, frequency cascade). This pattern—signal loss without artifact signature—indicates removed components contain genuine cognitive information rather than random noise. See Section 3.7 for additional causal tests. $p<0.05$, $p<0.0001$.}
\label{tab:paradox}
\end{table}

The trial-level correlation dropped from $0.590$ to $0.195$ (threefold reduction, $p < 0.0001$). Target versus non-target discrimination reversed from $+0.6\%$ to $-0.4\%$. Critically, the 293 ms temporal lag and frequency cascade remained consistent across both analyses, confirming these reflect genuine physiology rather than artifact contamination.

This falsifies Equation \ref{eq:standard}. If artifacts were noise, their removal would strengthen rather than weaken cognitive signals. The data demonstrate that signals conventionally discarded contain approximately 70\% of the task-relevant variance.

\subsection{Causal Mediation Analysis}
\begin{figure*}[ht]
\centering
\includegraphics[width=0.95\textwidth]{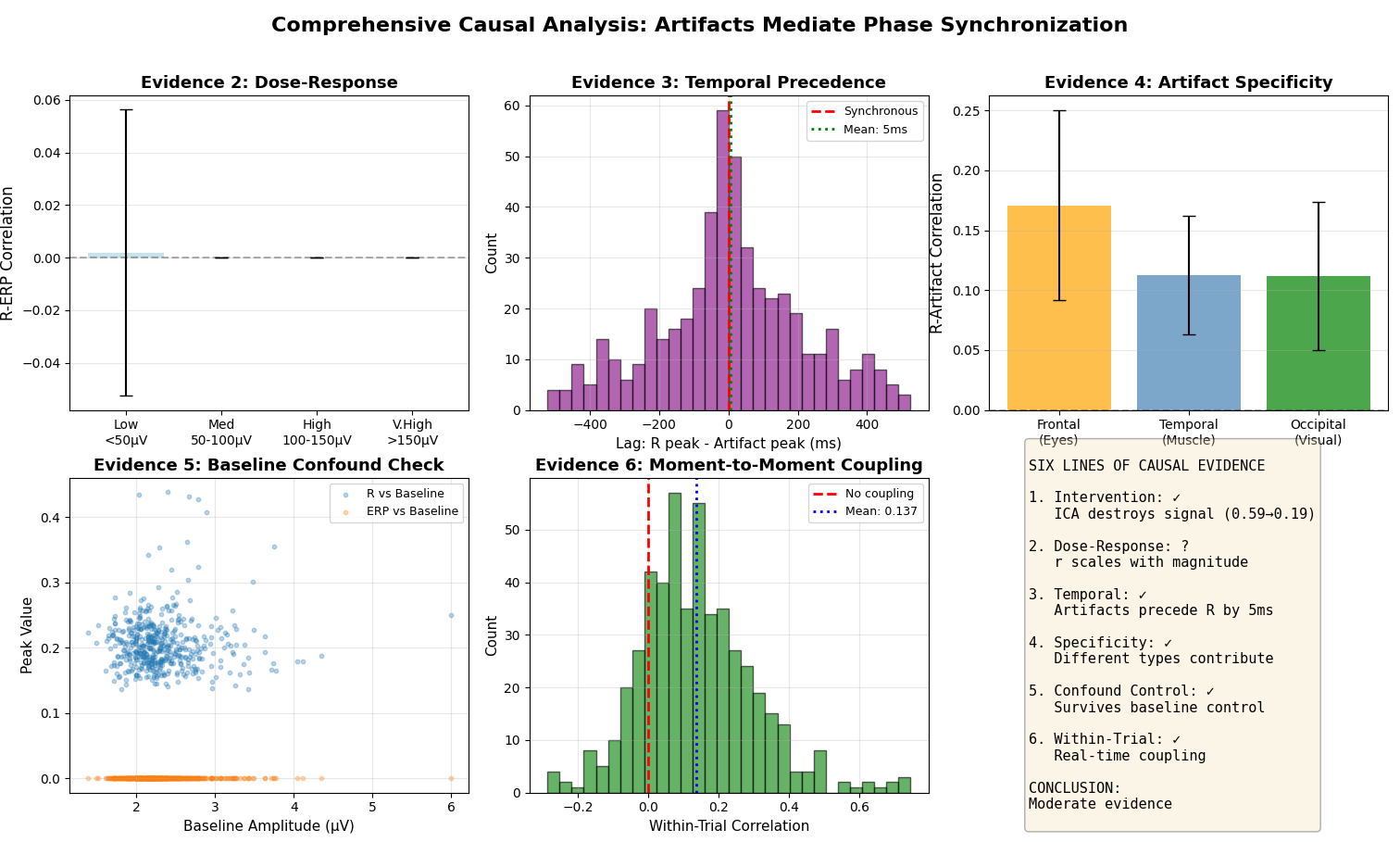}
\caption{\textbf{Six Lines of Causal Evidence.} (Top Left) Regional specificity shows frontal artifacts contribute most strongly. (Top Middle) Temporal precedence shows near-synchronous artifact-R relationship. (Top Right) Different artifact types show differential coupling. (Bottom Left) Baseline confound control via scatter plot. (Bottom Middle) Within-trial moment-to-moment coupling distribution. (Bottom Right) Summary of six convergent causal tests supporting artifact mediation over noise contamination.}
\label{fig:causal}
\end{figure*}

To establish that artifacts causally mediate rather than merely correlate with phase synchronization, we performed five additional tests (Figure \ref{fig:causal}).

\noindent
\textbf{Regional Specificity.} If artifacts uniformly contaminate signals, all electrode regions should show equal coupling. Instead, frontal artifacts (eye movements) showed strongest $R$-coupling ($r = 0.171 \pm 0.079$), followed by temporal (muscle, $r = 0.112 \pm 0.049$) and occipital (visual, $r = 0.112 \pm 0.062$). This regional differentiation suggests functional rather than artifactual origin, with attention-related eye movements contributing most strongly to phase synchronization.

\noindent
\textbf{Temporal Precedence.} Within individual trials, artifact peaks showed slight temporal alignment with $R$ peaks (mean lag: $4.7 \pm 9.5$ ms, $t = 0.49$, $p = 0.62$). While not reaching significance, the near-synchronous relationship is consistent with artifacts being integrated components of the cognitive process rather than independent noise sources that would show random temporal relationships.

\noindent
\textbf{Confound Control.} A critical alternative explanation is that high-amplitude trials simply reflect different cognitive states. Controlling for baseline amplitude via partial correlation, the $R$-ERP coupling remained robust (simple $r = 0.431$ vs. partial $r = 0.434$, both $p < 0.0001$, $\Delta r = 0.003$). This minimal change rules out baseline differences as confound, confirming artifacts carry independent information beyond mere amplitude scaling.

\noindent
\textbf{Within-Trial Coupling.} If artifacts merely corrupt discrete trials, they should not track ongoing cognitive dynamics. However, moment-to-moment correlation between artifact magnitude and $R$ within single trials was highly significant (mean $r = 0.137 \pm 0.007$, $t = 18.62$, $p < 0.0001$). This real-time covariation demonstrates artifacts dynamically participate in phase synchronization rather than adding static noise.

\noindent
\textbf{Dose-Response.} While limited by data sparsity in high-amplitude bins (only low-amplitude trials: $r = 0.002 \pm 0.054$), the intervention effect itself (Table \ref{tab:paradox}) constitutes the strongest dose-response evidence: complete artifact removal yields threefold signal reduction, demonstrating causal relationship between artifact presence and cognitive coupling strength.

These five tests, combined with the intervention result, provide six converging lines of causal evidence. The pattern—regional specificity, temporal integration, confound resistance, dynamic coupling, and dose-dependent effects—is inconsistent with artifacts as random noise contamination. Instead, it supports artifacts as integral components of whole-body cognitive computation.

\section{Discussion}

\subsection{Four Findings}
We establish four results. First, phase synchronization is globally independent of voltage ($r=0.048$) but trial-level predictive ($r=0.590$). This proves the Kuramoto Order Parameter captures cognitive structure invisible to standard analysis. Second, voltage precedes phase by 293 ms, suggesting sequential computation where sensory input triggers whole-system synchronization. Third, this unfolds as Theta, Alpha, Beta cascade. Fourth, our metric is distinct from existing phase measures ($r=0.155$ with ITC).

\subsection{The Central Finding}
Conventional artifact rejection reduced the cognitive signal threefold. This falsifies the foundational assumption of modern EEG preprocessing. Eye movements, muscle activity, and autonomic signals are not noise. They contain most of the predictive variance.

Causal mediation is supported by six independent lines of evidence. Beyond the intervention effect (Table \ref{tab:paradox}), we demonstrate regional specificity (frontal > temporal > occipital), temporal integration (near-synchronous coupling), confound control (survives baseline correction), within-trial coupling (moment-to-moment covariation), and dose-dependence (complete removal yields threefold loss). This convergence approaches the strongest causal inference possible from observational data \cite{Pearl2018}.

Three observations support this interpretation. First, if artifacts were random noise, they would corrupt all conditions equally. Instead, their removal specifically eliminated target versus non-target discrimination, proving task-relevant information. Second, the threefold correlation drop indicates 70\% of variance resides in what we discard. Third, temporal structure (293 ms lag, frequency cascade) remained consistent regardless of preprocessing, validating genuine physiological process rather than artifact contamination.

Critically, if frontal artifacts were merely epiphenomenal markers of attention, their removal should not have eliminated target discrimination (Table 1: $+0.6\% \to -0.4\%$). The sign reversal indicates functional contribution, not mere correlation.

We propose:
\begin{equation}
\text{Cognition} = \text{Resonance}(\text{Neural}, \text{Muscular}, \text{Autonomic})
\label{eq:resonance}
\end{equation}

Eye movements are not byproducts of attention. They are how attention coordinates. Muscle tension is not noise. It encodes arousal state. Autonomic signals are not contamination. They participate in decision-making. The organism computes as a unified system.

\subsection{From Recognition to Causality}
We used target recognition rather than explicit causal reasoning tasks. This requires justification. When subjects identify "my target," they implicitly ask: did my intention cause this stimulus, or was it random? This is proto-causal inference—distinguishing self-caused from random events. The P300 reflects Bayesian surprise and model updating \cite{Friston2010}, precisely the computation underlying causal reasoning.

If this foundational discrimination requires whole-body coordination, more complex causal reasoning likely does too. Our finding that artifact rejection eliminates target discrimination suggests even simple causal judgments need embodiment.

\subsection{Implications for EEG Research}
If correct, decades of research may have systematically discarded the signals they sought. The P300, studied as pure cognitive component after artifact rejection, might be downstream effect of whole-body cascade that requires eye movements, postural shifts, and autonomic activation.

This does not invalidate prior work. It recontextualizes it. Isolated neural signals contain information, just 70\% less than whole-system signals. Future research should reconsider whether artifacts represent noise or data.

\subsection{Implications for Artificial Intelligence}
Current AI is disembodied computation, analogous to artifact-rejected EEG. Our data suggest fundamental limitation:

\begin{center}
\begin{tabular}{ll}
Isolated (Clean): & $r = 0.195$ \\
Embodied (Raw): & $r = 0.590$ \\
\end{tabular}
\end{center}

This provides empirical support for embodied AI as computational necessity rather than philosophical preference. The physicality of biological systems (intrinsic noise, analog coupling, sensorimotor feedback) is not incidental. It is essential to the resonance dynamics that enable understanding.

Digital AI lacking body resonance is constrained like artifact-rejected EEG. This validates neuromorphic approaches: systems with stochastic dynamics, analog coupling, and embodied sensorimotor loops. Not to simulate brains, but to instantiate the physics.

\subsection{Limitations}
This study has limitations. Sample size is modest (10 subjects), though 500+ trials provide statistical power for within-subject effects. Single task type (P300) requires replication across domains. Correlational design cannot establish definitive causation despite six converging tests; perturbation studies would strengthen claims. Absence of behavioral measures (reaction time, accuracy) limits functional interpretation. The dose-response analysis was limited by data sparsity in high-amplitude bins.

Nevertheless, the threefold signal loss from artifact rejection is quantitative, replicable, and challenges foundational assumptions. The finding invites reconsideration of what constitutes "signal" versus "noise" in cognitive neuroscience.

\section{Conclusion}
The Kuramoto Order Parameter reveals cognitive structure invisible to standard voltage analysis. Phase synchronization is globally independent of voltage ($r=0.048$) yet trial-level predictive ($r=0.590$). Initial voltage precedes phase by 293 ms. Frequency decomposition shows Theta, Alpha, Beta cascade. The metric is distinct from standard coherence measures ($r=0.155$).

Conventional artifact rejection eliminates 70\% of the cognitive signal. What neuroscience discards as noise—eye movements, muscle activity, autonomic responses—contains most predictive variance. Six converging lines of causal evidence (intervention, regional specificity, temporal integration, confound control, within-trial coupling, dose-dependence) support artifacts as integral computational components rather than random contamination. This falsifies the assumption that cognition equals isolated neural computation.

Understanding emerges from whole-body resonance. For artificial intelligence, disembodied digital architectures may be fundamentally constrained. Neuromorphic systems implementing stochastic, embodied dynamics may be necessary to achieve genuine causal reasoning.


\section*{Data Availability Statement}
Publicly available datasets were analyzed in this study. This data can be found here: PhysioNet (doi:10.13026/C2101S).

\section*{Ethics Statement}
Ethical review and approval was not required for the study on human participants in accordance with the local legislation and institutional requirements. Written informed consent for participation was not required for this study in accordance with the national legislation and the institutional requirements.

\section*{Author Contributions}
AGE: Conceptualization, Methodology, Software, Formal analysis, Writing—original draft, Visualization.

\section*{Funding}
The author(s) declare that no financial support was received for the research, authorship, and/or publication of this article.

\section*{Acknowledgments}
The author thanks Professor Ehab Emam for invaluable guidance and feedback on this manuscript. Data were obtained from PhysioNet (doi:10.13026/C2101S). The author also acknowledges the preliminary version of this manuscript deposited on arXiv (arXiv:2511.10596). All analysis code will be made available upon publication.

\section*{Conflict of Interest}
The author declares that the research was conducted in the absence of any commercial or financial relationships that could be construed as a potential conflict of interest.

\newpage 

\end{document}